\documentclass[aps,prb,twocolumn,groupedaddress,showpacs]{revtex4}
\usepackage{amssymb}
\usepackage{graphicx}
\usepackage{amsmath}

\begin{document}

\title{Nanomechanical effects in~an~Andreev quantum dot}

\author{I.A.~Sadovskyy$^{a,b}$, G.B.~Lesovik$^{a}$, T.~Jonckheere$^{c}$, and T.~Martin$^{c,d}$}
\affiliation{$^{a}$L.D.~Landau Institute for Theoretical Physics RAS, Akad. Semenova av., 1-A, 142432, Chernogolovka, Moscow Region, Russia}
\affiliation{$^{b}$Moscow Institute of Physics and Technology, Institutskii per., 5, 141700, Dolgoprudny, Russia}
\affiliation{$^{c}$Centre de Physique Th\'eorique, Case 907 Luminy, 13288 Marseille Cedex 9, France}
\affiliation{$^{d}$Universit\'e de la M\'edit\'erann\'ee, 13288 Marseille Cedex 9, France}

\date{\today}

\begin{abstract}
We consider a quantum dot with mechanical degrees of freedom which is coupled to superconducting electrodes. A Josephson current is generated by applying a phase difference. In the absence of coupling to vibrations, this setup was previously proposed as a detector of magnetic flux and we wish here to address the effect of the phonon coupling to this detection scheme. We compute the charge on the quantum dot and determine its dependence on the phase difference in the presence of phonon coupling and Coulomb interaction. This allows to identify regions in parameter space  with the highest charge to phase sensitivity, which are relevant for flux detection. Further insight about the interplay of such couplings and subsequent entanglement properties between electron and phonon degrees of freedom are gained by computing the von Neumann entropy.
\end{abstract}

\pacs{
	74.45.+c,	% Proximity effects; Andreev effect; SN and SNS junctions
	73.21.La,	% Quantum dots
	74.78.Na,	% Mesoscopic and nanoscale systems
%	77.65.-j,	% Electromechanical resonance, 
	46.40.$-$f	% Mechanical vibrations, 
}

\maketitle

\section{Introduction}

The Josephson effect is one of the most striking manifestation of phase coherence in macroscopic objects. A nondissipative current~\cite{Josephson} can flow through a junction between two superconductors, provided that there is a phase difference between them. Early Josephson junctions consisted of an oxide layer or a normal metal sandwiched between the superconducting leads~\cite{Anderson_Rowell} but progress in nanofabrication techniques have allowed to imbed mesoscopic devices into the junction.~\cite{Park_2000,Pasupathy_2005,Kasumov_1999,cnt_JT,Bouchiat} One of the most commonly studied of such devices is the quantum dot. Quantum dots typically represent a normal metal island with resonant levels and possibly charging effects. In the context of Josephson transport, it has been shown theoretically that the charge on such quantum dots can be tuned either by applying a gate voltage to the dot or by varying the phase difference between the superconductors.~\cite{Engstrom_Kinaret,Sadovsky_Lesovik_Blatter_1} This continuous tuning of parameters allows the dot charge to deviate from an integer number. Of importance in such a system is that the tuning parameters can trigger a transition of the ground-state from a singlet (zero or double electron occupancy with opposite spins) to a doublet (single electron occupancy with spin up or spin down).~\cite{rozhkov_arovas,Sadovsky_Lesovik_Blatter_2} In Ref.~\onlinecite{Sadovsky_Lesovik_Blatter_2}, it has been proposed that the sensitivity of the dependence of the charge with respect to the flux could in principle be exploited to measure rather precisely the magnetic field in the loop, in the same spirit as a superconducting quantum interference device. The measurement of the charge itself could possibly be performed using a single-electron transistor coupled electrostatically to the dot in the junction.

At the same time, in nowadays experiments, one has the possibility either to taylor artificial quantum dots and to embed them in a circuit, or alternatively to use existing nano-objects for the same purpose. In particular, carbon nanotubes contacted to metallic or superconducting leads~\cite{Kasumov_1999,cnt_JT,Bouchiat} have been shown to behave like quantum dots, with the advantage that they can be influenced by nearby metallic gates.~\cite{cnt_JT,Bouchiat} There are also attempts to place single molecules in the junction between two reservoirs.~\cite{Park_2000,Pasupathy_2005} In such systems, the vibrational degrees of freedom may affect electron transport in two ways. First, there are always vibrational degrees of freedom associated with the material surrounding the molecular quantum dot. Such phonons typically constitute a source of relaxation and decoherence mechanism for quantum transport.\cite{Grange:2007,Jacak:2005,Ranninger:2002,Zhang:1999,Ridley:1982,Wilson-Rae:2002,Milde:2008,Kaer:2010,Weber:2010,Zazunov:2005,Ramsay:2010,Do:2007} Second, the quantum dot itself may have internal vibrational degrees of freedom, which are coupled to the charge of the quantum dot.\cite{Maier:2010,Tahir:2010,Zazunov:2010,JianXin:2003,Avriller:2009,Loos:2009,Haupt:2009,Yamamoto:2005} We focus on the latter mechanism in this work. A number of previous works have addressed this issue for nonequilibrium transport with normal metal contacts,~\cite{Jonsson_2005,Sapmaz_2006,Peng_2006,Witkamp,Ryndyk_Cuniberti,Tahir_MacKinnon,Huettel} or for the supercurrent through a vibrating nano-objects.~\cite{Zazunov,Novotny,Marchenkov1,Marchenkov2}

With this paper we want to address the issue of the phase sensitivity of the charge in an Andreev quantum dot, taking into account the presence of electron-phonon interaction. The goal is to determine the impact of the phonon coupling on the measurement scheme. Starting from a microscopic Hamiltonian model, we will compute the equilibrium properties of the system for various parameters, in the regime where the superconducting gap is much larger than all other relevant energies in the system.

\section{Model}

\begin{figure}[t]
  \includegraphics[width=8.3cm]{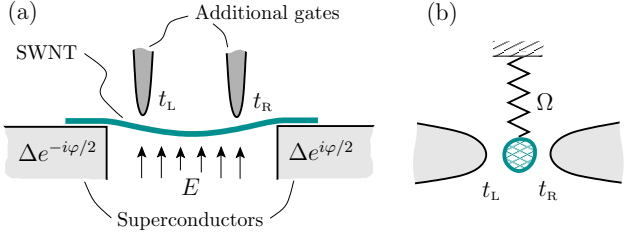}
  \caption[Nanomechanical setup]{
  (a) SWNT suspended between two superconducting leads with phase difference $\varphi$. The charge of the SWNT $Q$ is displaced by an external electrical field $E$. Two additional gates form the electron's resonances between them. (b) Model setup represents a quantum dot with oscillator degree of freedom coupled to superconductors through tunnel junctions.}
  \label{fig:nm-setup}
\end{figure}

Two typical setups are depicted in Fig.~\ref{fig:nm-setup}. On the right hand side is a generic setup where the central island is coupled to the right and left leads and which has a (single) vibrational degree of freedom. On the other hand, the  
setup on the left side [see Fig.~\ref{fig:nm-setup}(a)] of the figure represents a single-wall carbon nanotube (SWNT) which is suspended between two superconducting leads. Additional gates placed above the nanotube allow to define precisely the extent of the quantum dot, and therefore allow to modulate its energy levels. An overall gate voltage allows to apply an electric field to the whole structure. In this second setup, several vibrational modes are known to exist.~\cite{Marchenkov1,Marchenkov2}

We discuss primarily the case of a short junction (its length $L$ is much less than superconducting coherence length $\xi$) where the normal island can be described as a zero-dimensional (0D) object. Due to the shortness of the junction, in practice this setup effectively describes real 0D molecular quantum dots. Such setups were realized experimentally~\cite{Park_2000,Pasupathy_2005} and described theoretically.~\cite{Zazunov,Ryndyk_Cuniberti,Tahir_MacKinnon} 

The model of the Andreev quantum dot is described by a total Hamiltonian which includes the dot and its internal degrees of freedom, the leads, and the tunnel coupling between the latter two,
\begin{equation}
   {\hat H} = {\hat H}_{\rm\scriptscriptstyle D} + {\hat H}_{\rm\scriptscriptstyle S} + {\hat H}_{\rm\scriptscriptstyle T}.
   \label{H_tot}
\end{equation}
The first term ${\hat H}_{\rm\scriptscriptstyle D}$ is the quantum dot, which contains for simplicity a single level
and a discrete phonon spectrum,
\begin{multline}
   {\hat H}_{\rm\scriptscriptstyle D} = \sum\limits_i \hbar\Omega_i {\hat b}_i^\dag {\hat b}_i^{\phantom\dag} + \\
   \Big[ \varepsilon_{\rm\scriptscriptstyle D} - \sum\limits_i \lambda_i ({\hat b}_i^{\phantom\dag} + {\hat b}_i^\dag) \Big] \sum\limits_{\sigma=\uparrow, \downarrow} \Big({\hat n}_\sigma - \frac12 \Big) +
   U {\hat n}_\uparrow {\hat n}_\downarrow
   \label{H_D}
\end{multline}
with ${\hat n}_\sigma = {\hat d}_\sigma^\dag {\hat d}_\sigma^{\phantom\dag}$; ${\hat d}_\sigma^\dag$,  ${\hat d}_\sigma^{\phantom\dag}$ are electron creation and annihilation operators for the dot. The energy $\varepsilon_{\rm\scriptscriptstyle D}$ is the dot level, which can be tuned by a gate voltage, and which is measured with respect to the Fermi energy of the leads. Note that this simple model can represent a more realistic multi-level dot, when  one level only contributes significantly to the electronic transport because the spacing between the dot levels is large compared to the superconducting gap and the coupling to phonons. Each term $\hbar\Omega_i {\hat b}_i^\dag {\hat b}_i^{\phantom\dag}$ denotes the phonon energy of the vibration mode $i$ in the dot (${\hat b}_i^\dag$ and ${\hat b}_i^{\phantom\dag}$ are phonon creation and annihilation operators), $\lambda_i$ is the electron-phonon coupling in this mode; index $i$ runs over all mechanical modes $i=1,2, \ldots, N_{\rm modes}$. The electron-phonon coupling mechanism is described by terms ${\hat x}_i E$, where   ${\hat x}_i = \sqrt{\hbar/2M_i\Omega_i} ( {\hat b}_i^{\phantom\dag} + {\hat b}_i^\dag)$ are displacements in an external electrical field $E$. The charge of the dot is attracted by external gate voltage, which leads to a change in its position. The deformation leads to the changing of the ground-state energy and, therefore, of the charge of the dot. 

In the sum $\sum_\sigma({\hat n}_\sigma - 1/2)$ the constant $1/2$ is subtracted to ``symmetrize'' the matrix elements of the Hamiltonian. $U$ describes the Coulomb interaction. The lead Hamiltonian describes two BCS superconductors [with a lead index $\ell =\rm L, R$ (left, right)],
\begin{equation}
   {\hat H}_{\rm\scriptscriptstyle S} = \sum\limits_{\ell,k} {\hat \Psi}^\dag_{\ell,k}(\xi_k{\hat \sigma}_z + \Delta{\hat \sigma}_x) {\hat \Psi}^{\phantom\dag}_{\ell,k}, \quad
   {\hat \Psi}_{\ell,k} = \left[\!\! \begin{array}{l}
      \psi^{\phantom\dag}_{\ell,k,\uparrow} \\
      \psi^\dag_{\ell,-k,\downarrow}
   \end{array} \!\!\right]
   \label{H_S}
\end{equation}
with an energy dispersion in superconducting leads $\xi_k = \hbar^2 k^2/2m - \varepsilon_{\rm\scriptscriptstyle F}$ and an absolute value of the gap $\Delta$ in the bulk of the superconductors. The electron hopping term between dots and leads reads,
\begin{equation}
   {\hat H}_{\rm\scriptscriptstyle T} = \sum\limits_{\ell,k} \big( {\hat \Psi}^\dag_{\ell,k} \mathcal{\hat T}_\ell {\hat d} + {\rm h.c.} \big), \quad
   {\hat d} = \left[\! \begin{array}{l}
      {\hat d}_\uparrow \\
      {\hat d}_\downarrow^\dag
   \end{array} \!\right] \!,
   \label{H_T}
\end{equation}
where $\mathcal{\hat T}_{\rm\scriptscriptstyle L,R} = t_{\rm\scriptscriptstyle L,R} {\hat \sigma}_z e^{\pm i{\hat \sigma}_z \varphi/4}$ and $t_\ell$'s are tunneling amplitudes between superconductors and the dot. $\varphi$ is a superconducting phase difference. 

Calculations of observables for this system in thermal equilibrium typically start from the calculation of the partition function $Z \equiv {\rm Tr}\{ \exp (-\beta H) \}$, where $\beta \equiv 1/k_{\rm\scriptscriptstyle B}T$ is the inverse temperature. The Josephson current is then proportional to the logarithmic derivative with respect to the phase difference $\varphi$, and the charge on the dot is the derivative of the free energy with respect to the level position. In previous works using functional integral approaches~\cite{rozhkov_arovas} it was noted that because the total Hamiltonian is quadratic in the lead fermion operators, a partial trace over such degrees of freedom could be performed. This gives rise to an effective action with a dot fermion self-energy which contains retardation effects, and which couples fermion operators of the same nature, but with opposite spins. This coupling is a manifestation of electron pairing phenomena at the level of the dot due to the proximity with the superconducting leads. In Ref.~\onlinecite{Zazunov}, the calculation of the partial trace over the leads of the partition function was performed in a similar manner, nevertheless using an operator approach. Furthermore, the assumption that $|\varepsilon_{\rm\scriptscriptstyle D}|,$ $U,$ $\Gamma$, $\hbar\Omega \ll \Delta$ (the so-called $\Delta \to \infty$ limit) allowed there to neglect the retardation effect and to therefore derive an effective Hamiltonian for the dot-phonon system,
\begin{multline}
   {\hat H} = \sum\limits_i \hbar\Omega_i {\hat b}_i^\dag {\hat b}_i^{\phantom\dag} +
   \Big[ \varepsilon_{\rm\scriptscriptstyle D} - \sum\limits_i \lambda_i ({\hat b}_i^{\phantom\dag} + {\hat b}_i^\dag) \Big] \sum\limits_{\sigma=\uparrow, \downarrow} \Big({\hat n}_\sigma - \frac12 \Big) + \\
   + \tilde\Gamma [{\hat d}_\downarrow {\hat d}_\uparrow + {\rm h.c.}] +
   U {\hat n}_\uparrow {\hat n}_\downarrow,
   \label{H_eff}
\end{multline}
where $\tilde\Gamma = \Gamma \cos(\varphi/2)$. The escape rate $\Gamma = 2\pi \nu(0) |t|^2$ (or resonance width of the dot), assuming a constant density of states $\nu(0)$ near the Fermi energy of the metal in the normal state. We assume a symmetric setup $|t_{\rm\scriptscriptstyle L}|^2 = |t_{\rm\scriptscriptstyle R}|^2 = |t|^2$ for the remainder of this study. The effective Hamiltonian Eq.~(\ref{H_eff}) of the large $\Delta$ limit constitutes the starting point of our calculation.

The matrix elements of the Hamiltonian with respect to the dot electron states $|\nu\rangle_{\rm el}$ are now computed. These states are:
$|0\rangle_{\rm el}$ (zero occupation), 
$|\!\!\uparrow \rangle_{\rm el} \equiv {\hat d}_\uparrow^\dag |0\rangle_{\rm el}$, 
$|\!\!\downarrow \rangle_{\rm el}$ $\equiv {\hat d}_\downarrow^\dag |0\rangle_{\rm el}$ (single occupation), 
$|2\rangle_{\rm el} \equiv {\hat d}_\uparrow^\dag {\hat d}_\downarrow^\dag |0\rangle_{\rm el}$ 
(double occupation), which means that from the electron point of view, the only off-diagonal part of $\hat H$ originates from the coupling to the leads and involve either zero or double occupancy states,
\begin{multline}
   H_{\mu\nu}
   = \sum\limits_i \hbar\Omega_i {\hat b}_i^\dag {\hat b}_i^{\phantom\dag} +
   \Big[ \varepsilon_{\rm\scriptscriptstyle D} - \sum\limits_i \lambda_i ({\hat b}_i^{\phantom\dag} + {\hat b}_i^\dag) \Big] \times \\  {\rm diag}\{-1, 0, 0, 1\}
   + \tilde\Gamma
\left[ \!
\begin{array}{cccc}
 0 & 0 & 0 & 1  \\
 0 & 0 & 0 & 0  \\
 0 & 0 & 0 & 0  \\
 1 & 0 & 0 & 0  
\end{array}
\! \right]
 + U \, {\rm diag}\{0, 0, 0, 1\}.
   \label{H_eff_me}
\end{multline}
The basis of phonon states is
$|n\rangle_{{\rm ph},i} \equiv ({\hat b}_i^\dag)^n (n!)^{-1/2} \times$ $|0\rangle_{{\rm ph},i}$. % 
Using relations 
$_{i,{\rm ph}}\langle m|{\hat b}_i^\dag {\hat b}_i^{\phantom\dag}|n\rangle_{{\rm ph},i} = n\delta_{mn}$ and
$_{i,{\rm ph}}\langle m|{\hat b}_i^{\phantom\dag} + {\hat b}_i^\dag|n\rangle_{{\rm ph},i} = \sqrt{n} \, \delta_{m,n-1} + \sqrt{n+1} \, \delta_{m,n+1}$,
we can thus generate the phonon matrix elements of Eq.~(\ref{H_eff}). For one single phonon 
mode the matrix representing the full Hamiltonian reads,
\begin{equation}
   H_{\mu\nu, mn} =
\left[ \!
\begin{array}{ccccl}
 \Theta_1 & \Lambda_1 & 0 & 0 & \cdots \\
 \Lambda_1 & \Theta_2 & \Lambda_2 & 0 & \\
 0 & \Lambda_1 & \Theta_3 & \Lambda_3 & \\
 0 & 0 & \Lambda_3 & \Theta_4 & \\
 \vdots & & & & \ddots
\end{array}
\! \right] \!\!.
   \label{H_eff_me2}
\end{equation}
We use greek indices for matrix elements in electron subspace and latin ones for phonons. The $4 \times 4$ matrix blocks $\Theta_n$ and $\Lambda_n$ are defined by
\begin{equation}
   \Theta_{\mu\nu, n} = 
   \left[ \!
\begin{array}{cccc}
 -\varepsilon_{\rm\scriptscriptstyle D} & 0 & 0 & {\tilde\Gamma} \\
 0 & 0 & 0 & 0 \\
 0 & 0 & 0 & 0 \\
 {\tilde\Gamma} & 0 & 0 & \varepsilon_{\rm\scriptscriptstyle D} + U
\end{array}
\! \right]
   + n\hbar\Omega
   \label{H_el}
\end{equation}
and $\Lambda_{\mu\nu, n} = {\rm diag}\{-1, 0, 0, 1\} \sqrt{n}\lambda$; they describe electron degrees of freedom with $n$ phonons and electron-phonon coupling, respectively. The generalization to an arbitrary number of phonon modes can easily be obtained by multiplication of Hilbert subspaces for each phonon modes.

The eigenstates of the effective Hamiltonian can be calculated numerically by truncation of the matrix (truncation of the number of phonon states). In practice, we took about 20 phonon states for $\lambda/\Omega = 3$ and about 70 states for $\lambda/\Omega = 5$; these numbers are nearly independent of the number of modes $N_{\rm modes}$.

\section{States without phonons: $\lambda = 0$}

The electron states case were described using electron representation~\cite{rozhkov_arovas} and electron-hole Bogoliubov superposition.~\cite{Sadovsky_Lesovik_Blatter_2} The effective Hamiltonian reduces to $4 \times 4$ matrix $\Theta_0$ [see Eq.~(\ref{H_el})] with singly degenerated (singlets) eigenstates
\begin{equation}
   E_{0,2} = U/2 \mp \sqrt{(\varepsilon_{\rm\scriptscriptstyle D} + U/2)^2 + {\tilde\Gamma}^2}
   \label{S_en}
\end{equation}
and doubly degenerated (doublet) eigenstate
\begin{equation}
   E_1 \equiv E_{\uparrow,\downarrow} = 0.
   \label{D_en}
\end{equation}
In the absence of Coulomb interaction $U=0$ the eigenvalues are ordered as $E_0 < E_1 < E_2$ and the ground-state is always formed by the state with energy $E_0$ (pure hole-like state in Bogoliubov representation). For $U>0$ the ground-state can be formed by the singlet $|0\rangle_{\rm el}$ [the singlet region in 2D plane ($\varepsilon_{\rm\scriptscriptstyle D}$, $\varphi$)] or by the doublet $|1\rangle_{\rm el}$ (doublet region), but never by $|2\rangle_{\rm el}$. For arbitrary finite $U$ the doublet region exists if
\begin{equation}
   (\varepsilon_{\rm\scriptscriptstyle D} + U/2)^2 + {\tilde\Gamma}^2 < (U/2)^2,
   \label{D_reg}
\end{equation}
where $\tilde\Gamma = \Gamma \cos(\varphi/2)$.

\section{Regions of the singlet and doublet states}

\begin{figure}[b]
  \includegraphics[width=8.6cm]{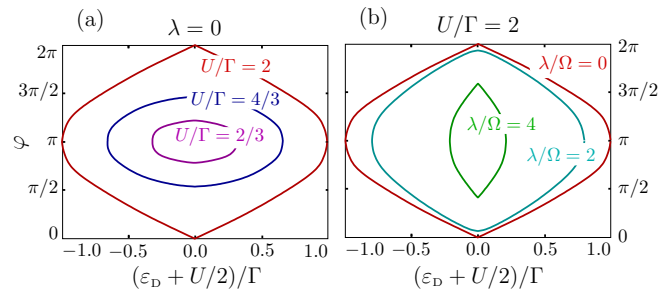}
  \caption[Doublet region]{
  Singlet-doublet jump in the $(\varphi, \varepsilon_{\rm\scriptscriptstyle D})$ plane; the singlet regions lie outside the color loops, the doublet regions lies inside the loops. 
  (a) For zero electron-phonon interaction $\lambda=0$ and different Coulomb interaction $U/\Gamma = 0$ (empty region), $2/3$ (magenta; smallest region), $4/3$ (blue), $2$ (red; largest region). It is described by Eq.~(\ref{D_reg}).
  (b) For different phonon-electron interaction strength $\lambda = 0$ (red; largest region), $\lambda =2$ (cyan), $\lambda =4$ (green; smallest region) and fixed Coulomb interaction $U/\Gamma=2$. The single-mode oscillator has a frequency which is much smaller than the tunnel resonance width $\hbar\Omega/\Gamma=0.05$. We see the interplay between Coulomb interaction and electron-phonon coupling; the first one expands the doublet region in the $\varphi$ direction, the second one squeezes it in the $\varepsilon_{\rm\scriptscriptstyle D}$ direction.} 
  \label{fig:D_reg}
\end{figure}

The existence of the doublet state as the ground-state of the system is important for this system and we dwell on this more. For zero electron-phonon interaction $\lambda=0$ the doublet region is specified by Eq.~(\ref{D_reg}) and its form is represented in Fig.~\ref{fig:D_reg}(a). The highest value of $U$ corresponds to the largest doublet region; with decreasing of $U$ this region becomes smaller and smaller. At $U=0$ its disappears. 

If we start with some fixed finite $U$ then the ``area'' of the doublet region decreases with electron-phonon coupling $\lambda$. The evolution of the doublet region is plotted in Fig.~\ref{fig:D_reg}(b) for different values of the electron-phonon coupling constant. The largest loop corresponds to the smallest (zero) $\lambda$. Upon switching $\lambda$, the reduction of this region is barely noticeable, it acts mostly on the level position range as it still approaches the phase values $0$ and $2\pi$. There is a competition between charge repulsion effects on the dot and the presence of the electron-phonon coupling, which can be understood to be playing the role of an effective attractive, retarded, interaction. This explains the reduction of the doublet region. 

At horizontal line $\varphi=\pi$ the decreasing of the doublet region in $\varepsilon_{\rm\scriptscriptstyle D}$ direction can be described by inequation $|\varepsilon_{\rm\scriptscriptstyle D} + U/2| < U/2 - \lambda^2/\hbar\Omega$. The nonzero electron-phonon coupling acts as the negative Coulomb interaction (in the sense of size of the doublet region). It implies that for all $\lambda$'s larger than 
\begin{equation}
   \lambda_{\rm\scriptscriptstyle C} = \sqrt{\hbar\Omega U/2}
   \label{lambda_crit}
\end{equation}
the doublet region does not exist for any values of $\varepsilon_{\rm\scriptscriptstyle D}$ and $\varphi$.

\section{Charge of the Andreev dot}

The charge $Q$ of the nanotube/quantum dot [in a given quantum mechanical state, e.g., some eigenstate of~(\ref{H_eff})] can be calculated by taking the derivative of its energy $E$ (in this particular state) with respect to the external gate potential $V_{\rm g}$ (or $\varepsilon_{\rm\scriptscriptstyle D}/e$),
\begin{equation}
   Q = e \, \partial E / \partial \varepsilon_{\rm\scriptscriptstyle D}.
\end{equation}
If one is interested in the charge of the ground-state, then ground-state energy should be taken.

The same result can be obtained by averaging the charge operator
\begin{equation}
   {\hat Q} = e \sum\limits_{\sigma=\uparrow, \downarrow} \Big({\hat n}_\sigma - \frac12 \Big)
   \label{Q_op}
\end{equation}
and the corresponding matrix elements
\begin{equation}
   Q_{\mu\nu} = e \, {\rm diag}\{-1, 0, 0, 1\}
   \label{Q_me}
\end{equation}
over the needed state [Eq.~(\ref{Q_me}) is written in the electron subspace; it should be multiplied by the unity matrix in the phonons subspace]. In what follows we concentrate on the behavior of the charge as a function of flux $\varphi$ and dot level position $\varepsilon_{\rm\scriptscriptstyle D}$.

\begin{figure}[t]
  \includegraphics[width=8.4cm]{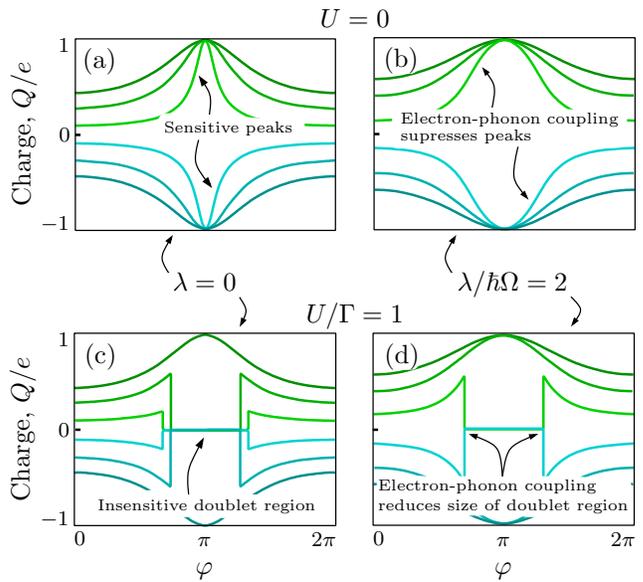}
  \caption[Charge of the Andreev dot as a function of $\varphi$]{
Charge as a function of $\varphi$ at different values of $\varepsilon_{\rm\scriptscriptstyle D}$ (from top to bottom $\varepsilon_{\rm\scriptscriptstyle D}/\Gamma = -0.5$, $-0.3$, $-0.1$, $0.1$, $0.3$, $0.5$) for zero Coulomb interaction $U=0$ and different $\lambda$'s. Phonon frequency $\Omega$ is much smaller than tunnel resonance width $\Gamma$, $\hbar\Omega/\Gamma = 0.05$.
  (a) and (b) Zero Coulomb interaction $U = 0$. $\lambda=0$ and the charge $Q_0$ is given by Eq.~(\ref{S_Q}) at (a). $\lambda/\hbar\Omega=2$ at (b). The maximum values of the differential charge-to-flux sensitivity [Eq.~(\ref{Sens})] is near the point $\varphi = \pi$ with parameter $\varepsilon_{\rm\scriptscriptstyle D}$ around zero. For the symmetric barrier and $U=0$ the sensitivity $\mathcal{S}(\varphi, \varepsilon_{\rm\scriptscriptstyle D})$ has the singularity at $(\pi,0)$ which disappears with any finite asymmetry, Coulomb interaction, or electron-phonon interaction.
  (c) and (d) The same for finite but small Coulomb interaction $U = \Gamma$. The ``flat'' doublet region appears. The width of this region decreases as $\lambda$ increases: from (c) to the (d) plot.
  }
  \label{fig:Charge}
\end{figure}
In the absence of phonons ($\lambda=0$) the dot charge can be found from Eqs.~(\ref{S_en}) and (\ref{D_en}). For the case when the singlet constitutes the ground-state
\begin{equation}
   Q_{0,2} = \mp e\frac{
       \varepsilon_{\rm\scriptscriptstyle D} + U/2
   }{
       \sqrt{(\varepsilon_{\rm\scriptscriptstyle D} + U/2)^2 + {\tilde\Gamma}^2}
   };
   \label{S_Q}
\end{equation}
[we should add the electron charge to this result if we remember about the subtraction $\sum_\sigma 1/2 = 1$ which appears in the dot Hamiltonian~(\ref{H_D})]. For the case of the doublet we find:
\begin{equation}
   Q_1 = 0.
   \label{D_Q}
\end{equation}
[or $Q_1 = e$ if we restore the constant term which is subtracted in the Hamiltonian~(\ref{H_D})]. Everywhere in this article we keep $Q_1 = 0$ for symmetry, but the unit charge is well defined (no quantum fluctuations) and has the correct physical interpretation. Let us start with a normal dot with some well-defined integer charge $q=0,e,2e$. Then we connect the superconductors through tunnel barriers to superconductors (with Cooper pairs). If the charge is odd $q=e$ then it does not ``feel'' the superconductors and the charge remains integer. In the case of even initial charge $q=0,2e$ it couples with Cooper pairs in superconductors and creates the singlet state with fractional and fluctuating charge (with rms value about $e$). 

For zero Coulomb interaction $U=0$ the doublet region is absent and for $\lambda=0$ the charge is given by $Q_2$ fom Eq.~(\ref{S_Q}) everywhere, see Fig.~\ref{fig:Charge}(a). Note, that the charge $Q_2$ near the values $(\varphi, \varepsilon_{\rm\scriptscriptstyle D}) = (\pi, 0)$ has a narrow peak and it changes its sign with $\varepsilon_{\rm\scriptscriptstyle D}$. For $U=\lambda=0$ this peak corresponds to an infinite ``charge-to-phase sensitivity,'' see Sec.~\ref{sec:Sensitivity}. For asymmetric barriers there is no point with infinite slope and the maxima of sensitivity are reached at two locations around $\varphi=\pi$, see Ref.~\onlinecite{Sadovsky_Lesovik_Blatter_2}. Note that this peak is broadened by temperature, finite superconducting gap $\Delta$, Coulomb interaction, and electron-phonon interaction. The later is shown in Fig.~\ref{fig:Charge}(b).

Next, if one now considers nonzero Coulomb interaction [Fig.~\ref{fig:Charge}(c)], then a ``flat'' doublet region exists: the infinitely narrow peak disappears for any $U>0$ and in the singlet region the charge is still given by Eq.~(\ref{D_Q}). 

In addition, at the boundary of the doublet and the singlet region the charge exhibits jumps (for the finite superconducting gap $\Delta$ or temperature $T>0$ this jump is smeared). Therefore the sensitivity is once again singular because one abruptly changes the nature of the ground-state upon varying $(\varphi, \varepsilon_{\rm\scriptscriptstyle D})$. 

Further we study the combination of the charging effects in the dot and the electron-phonon coupling. It turns out that the size of the doublet region decreases as the strength of the electron-phonon interaction $\lambda$ (more precisely, the factor $\lambda/\hbar\Omega$) increases. This is displayed in Figs.~\ref{fig:D_reg}(b), \ref{fig:Charge}(c) and \ref{fig:Charge}(d). Comparing the Figs.~\ref{fig:Charge}(c) and ~\ref{fig:Charge}(d) it can be seen that the overall topology of the plots is the same except for the fact that the reduced doublet region persists at $U\neq 0$.  

In the Appendix~\ref{app:Charge} we provide a more detailed explanation of the properties of the function $Q(\varphi, \varepsilon_{\rm\scriptscriptstyle D})$.

\section{Charge-to-phase sensitivity
\label{sec:Sensitivity}}

One can measure the charge via a capacitively connected charge detector, e.g., a single-electron transistor. It does not ``feel'' the whole Andreev quantum dot charge $Q$ in practice, but it feels some renormalized charge $\alpha_{\rm\scriptscriptstyle C} \alpha_{\rm\scriptscriptstyle S} Q$. The geometrical factor $\alpha_{\rm\scriptscriptstyle C}$ comes from the properties of the measurement gate (capacitance $C_{\rm m}$) and other capacitively connected parasitic things around (capacitance $C_{\rm o}$): $\alpha_{\rm\scriptscriptstyle C} = C_{\rm m} / (C_{\rm m} + C_{\rm o})$. The second factor $\alpha_{\rm\scriptscriptstyle S}$ comes from the dynamical feedback of the charge detector. In this article we suppose for simplicity that $\alpha_{\rm\scriptscriptstyle C} = \alpha_{\rm\scriptscriptstyle S} = 1$ keeping in mind that the charge in some way is suppressed during measurement procedure (we thus study the  charge sensitivity unperturbed by detector). 

\begin{figure}[t]
    \includegraphics[width=8.4cm]{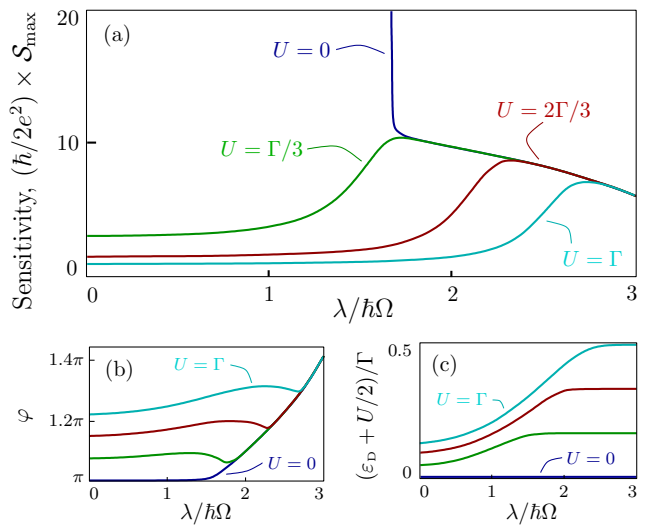}
  \caption[Charge-to-phase sensitivity]{
     (a) The maximal value of charge-to-phase sensitivity [see Eq.~(\ref{Sens})] at $(\varphi, \varepsilon_{\rm\scriptscriptstyle D})$ plane as a function of electron-phonon coupling strength $\lambda$ at different Coulomb energies $U$.
     (b) and (c) The values of $\varphi$ and $\varepsilon_{\rm\scriptscriptstyle D}$ at which the maximum of the sensitivity is attained.
     At all plots the $U=0$ (blue), $\Gamma/3$ (green), $2\Gamma/3$ (red), $\Gamma$ (cyan).
     The sensitivity can decreases or increases with $\lambda$. At $U=0$ the sensitivity goes from non-physical infinity value at $\lambda=0$; this infinity by asymmetry of the dot, finite of $\Delta$, temperature, etc. $\hbar\Omega/\Gamma = 0.05$.
  }
  \label{fig:Sensitivity}
\end{figure}
Let us define the charge-to-phase sensitivity at a given point as the derivative
\begin{equation}
   \mathcal{S} = \frac{2e}{\hbar} \frac{\partial Q}{\partial (\varphi/2)}.
   \label{Sens}
\end{equation}
This quantity characterizes the charge response to the superconducting phases difference, and, hence, to the magnetic flux. It can be useful for a flux a detector which is based on measuring the charge in the Andreev quantum dot.~\cite{Sadovsky_Lesovik_Blatter_2} Note that Eq.~(\ref{Sens}) coincides with the current-to-gate voltage sensitivity $\mathcal{S} = e \, \partial I / \partial \varepsilon_{\rm\scriptscriptstyle D}$ [$= (2e^2/\hbar) \, \partial^2 E / \partial \varepsilon_{\rm\scriptscriptstyle D} \partial (\varphi/2)$].

Consider the structure of the sensitivity as a function of the parameters $(\varphi, \varepsilon_{\rm\scriptscriptstyle D})$ and its maxima in these parameters $\mathcal{S}_{\rm max}$. In this article we concentrate on the sensitivity of the singlet region and we omit the sensitivity due to the jumps of the charge at the singlet-doublet border.

For $U=\lambda=0$ the sensitivity has a ``meaningless'' large value at $(\varphi, \varepsilon_{\rm\scriptscriptstyle D}) = (\pi, 0)$ which corresponds to the narrow peak in the charge, see Fig.~\ref{fig:Charge}(a). The interaction with the vibrating mode cuts this value and the sensitivity decreases with $\lambda$, which is shown in Fig.~\ref{fig:Sensitivity}(a) by the top line [the maximum moves away from the point $(\pi, 0)$ and its new position is shown in Figs.~\ref{fig:Sensitivity}(b) and \ref{fig:Sensitivity}(c)].

Given a finite $U$ the sensitivity initially is totally suppressed by the existence of the non-sensitive doublet region, whereas the maximum sensitivity moves to the border of the singlet and doublet regions. Increasing $\lambda$, the sensitivity of the singlet region goes down, but the size of the doublet region decreases. The competition between these two effects gives us new maxima --- lower lines in Fig.~\ref{fig:Sensitivity}(a). The effect of the decreasing doublet region ``wins'' when the curve goes up (small $\lambda$'s, the maximal sensitivity at the singlet-doublet border); when the size of the doublet region is small enough the sensitivity has its maximum inside the singlet region and does not depend on $U$ --- curves merge and go down.

\section{Entropy}

\begin{figure}[t]
  \includegraphics[width=7.6cm]{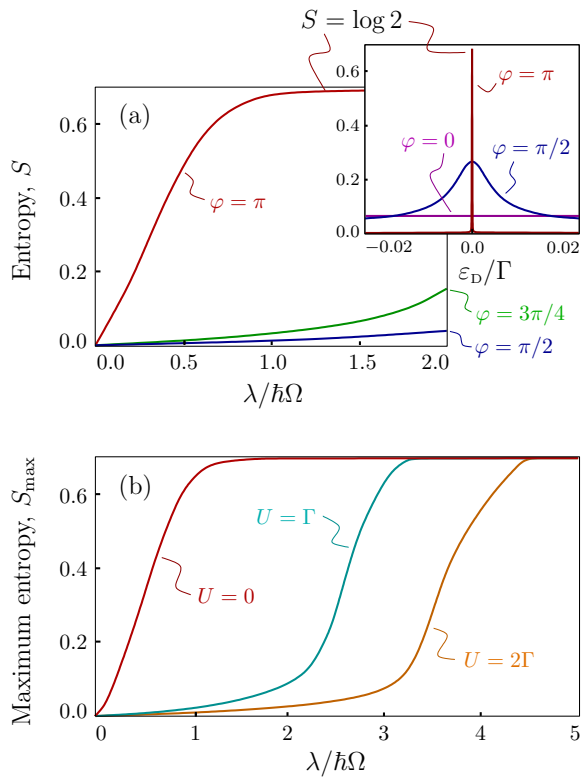}
  \caption[Entropy of the electron subsystem]{
(a) Subsystem entropy [Eq.~(\ref{Ent})] at $\varepsilon_{\rm\scriptscriptstyle D}=0$ and different $\varphi$ as a function of $\lambda$, with $\hbar\Omega/\Gamma = 0.05$ and $U=0$. The entropy increases from zero as the parameter $\lambda/\hbar\Omega$ increases, and saturates at value $S = \log 2$.
Top inset: dependence of the entropy on dot level $\varepsilon_{\rm\scriptscriptstyle D}$ at superconducting phase difference $\varphi = 0$ (no dependence), $\pi/2$ and $\pi$ (strong peak around $\varepsilon_{\rm\scriptscriptstyle D} = 0$) for phonon-electron interaction strength $\lambda/\hbar\Omega=3$.
(b) Maximum entropy in the $(\varphi, \varepsilon_{\rm\scriptscriptstyle D})$ space as a function of $\lambda/\hbar\Omega$ for $U=0$ (red), $U=\Gamma$ (cyan), and $U=2\Gamma$ (brown). The maximum
entropy is reached for $\varepsilon_{\rm\scriptscriptstyle D}=0$, and values of $\varphi$ which depend
on $U$.
  }
  \label{fig:Entropy}
\end{figure}
The entropy provides a measure of the effectiveness of the electron-phonon coupling, and in particular to what extend this coupling entangles the electron and phonon degrees of freedom. The density matrix of the total system is given by ${\hat \rho} = |\Psi \rangle \langle\Psi|$, where $|\Psi \rangle$ is the ground eigenstate of the Hamiltonian ${\hat H}$, see Eq.~(\ref{H_eff}). The density matrices of the electron and phonon subsystems are defined as ${\hat \rho}_{\rm el} = {\rm Tr}_{\rm ph} \{ {\hat \rho} \}$ and ${\hat \rho}_{\rm ph} = {\rm Tr}_{\rm el} \{ {\hat \rho} \}$, respectively (here ${\rm Tr}_{\rm ph}$ and ${\rm Tr}_{\rm el}$ denote traces over electron and phonon degrees of freedom). Given a subsystem density matrix, the von Neumann entropy is defined as
\begin{equation}
   S 
   = - {\rm Tr}\{{\hat \rho}_{\rm ph} \log {\hat \rho}_{\rm ph} \} 
   \equiv - {\rm Tr}\{{\hat \rho}_{\rm el} \log {\hat \rho}_{\rm el} \}.
   \label{Ent}
\end{equation}

We consider for simplicity a single phonon mode. Let us discuss first the absence of Coulomb interaction ($U=0$); then the system can never be in the doublet state, and the electronic basis can be restricted to $[|0\rangle_{\rm el}, |2 \rangle_{\rm el}]^{\rm T}$ and the ground-state can be written in full generality
\begin{equation}
   |\Psi \rangle = 
   a_0 \, |0 \rangle_{\rm el} \otimes |p_0\rangle_{\rm ph} + 
   a_2 \, |2 \rangle_{\rm el} \otimes |p_2\rangle_{\rm ph},
\end{equation}
where $|0\rangle_{\rm el}$ and $|2 \rangle_{\rm el}$ are the electronic states, $|p_0\rangle_{\rm ph}$ and $|p_2\rangle_{\rm ph}$ are normalized phonon states (which can be expressed as linear combinations of the basis phonon states $|n\rangle_{\rm ph}$), and $|a_0|^2 + |a_2|^2 = 1$. The reduced density matrix of the electron subsystem is then
\begin{equation}
   {\hat \rho}_{\rm el} = \left(
   \begin{array}{cc}
      |a_0|^2 & a_0^* a_2 \langle p_2 | p_0 \rangle \\
      a_0 a_2^* \langle p_0 | p_2 \rangle & |a_2|^2 
   \end{array} \right)
\end{equation}
Two extreme cases are simple and notable. First, for $\langle p_0|p_2\rangle=1$, the density matrix corresponds to a pure state with zero entropy; accordingly the wave function can be factorized as 
$|\Psi \rangle = (a_0 |0\rangle_{\rm el} + a_2 |2\rangle_{\rm el}) \otimes |p_0\rangle_{\rm ph}$. 
Second, when $\langle p_0 | p_2 \rangle =0$, then the density matrix is diagonal, with $S=-|a_0|^2 \log |a_0|^2 - |a_2|^2 \log |a_2|^2 $, which gives the maximum value $S=\log 2$ when $|a_0|^2 = |a_2|^2 = 1/2$. In the general case, the entropy is $S=-\rho_+ \log \rho_+ -\rho_- \log \rho_-$, with the eigenvalues $\rho_\pm = 1/2 \pm [1/4 - |a_0|^2 |a_2|^2 (1-|\langle p_0|p_2 \rangle|^2)]^{1/2}$. As entropy is maximal ($S=\log 2$) when $\rho_+ = \rho_- = 1/2$, and decreases as the difference between $\rho_+$ and $\rho_-$ increases, we see that to have a large entropy one needs to have $\langle p_0 | p_2 \rangle$ as small as possible, and $|a_0|^2 = 1 - |a_2|^2$ as close to $1/2$ as possible.

The behavior of the entropy as a function of the parameters $\lambda$ (phonon coupling), $\varepsilon_{\rm\scriptscriptstyle D}$ (position of the dot level) and ${\tilde\Gamma}=\Gamma \cos(\varphi/2)$ is shown in Fig.~\ref{fig:Entropy}(a).

The main panel of the figure shows the increase in the entropy as a function of $\lambda$, for different values of $\varphi$, and $\varepsilon_{\rm\scriptscriptstyle D}=0$. The fastest increase is obtained for $\varphi=\pi$ ($\tilde\Gamma=0$). 
The entanglement of electrons with the other degrees of freedom develops easier, if the two electronic levels cross or close to each other. The biggest entanglement/entropy at $\varphi = \pi$ then looks natural, since  the energies $E_-$ and $E_+$ [see Eq.~(\ref{S_en})] coincide at the point $(\pi, 0)$. The difference of $E_+ - E_-$ increases with increasing ``distance'' from point $(\varphi, \varepsilon_{\rm\scriptscriptstyle D})$ to point $(\pi, 0)$, and correspondingly away from the point $(\pi, 0)$ entanglement decreases. Also the  increasing  of the entropy with $\lambda$ can be understood from simple analysis of the nature of the electron-phonon coupling: the coupling to the electronic levels $|0\rangle_{\rm el}$ and $|2 \rangle_{\rm el}$ ``displaces'' the phonon field in opposite directions, thus making the phonon states overlap $|\langle p_0 | p_2 \rangle|$ smaller as $\lambda$ increases, which increases entropy. Decreasing $\varphi$ (thus increasing $\tilde\Gamma$) gives a smaller entropy. This is due to the coupling between the states $|0\rangle_{\rm el}$ and $|2 \rangle_{\rm el}$ when ${\tilde\Gamma}\neq 0$; the states $|p_0\rangle$ and $|p_2\rangle$ are then combinations of the two displaced states, which makes the overlap $|\langle p_0 | p_2 \rangle|$ larger and thus decreases entropy.  

The inset of the figure shows how the entropy varies when $\varepsilon_{\rm\scriptscriptstyle D}$ is changed, for different values of $\varphi$: it has a peaked behavior, with a width which decreases sharply as $\varphi$ gets closer to $\pi$ (that is, $\tilde\Gamma$ to 0); for $\varphi=\pi$, the width of the peak is precisely zero: since the electronic levels $|0\rangle_{\rm el}$ and $|2\rangle_{\rm el}$ are not coupled for $\varphi=\pi$, any non-zero value of $\varepsilon_{\rm\scriptscriptstyle D}$ means that the ground-state is obtained with a single electronic state only ($|0\rangle_{\rm el}$ or $|2\rangle_{\rm el}$), and thus entanglement with the phonon field, and entropy, is zero.  

Let us now consider the effect of Coulomb interaction ($U > 0$). As has been shown in previous sections, it creates a doublet region. There, the entropy is simply zero. When a doublet region exists, the maximal entanglement between electron and mechanical subsystems is achieved at the border of the singlet/doublet regions. Therefore the maximum entropy (in variables $(\varphi, \varepsilon_{\rm\scriptscriptstyle D})$) is obtained for $\varepsilon_{\rm\scriptscriptstyle D}=0$ and $\varphi$ at the edge of the singlet region. The entropy as a function of $\lambda$ for non-zero $U$ is plotted in Fig.~\ref{fig:Entropy}(b). The red curve ($U=0$) is the same as the red curve in Fig.~\ref{fig:Entropy}(a); for $U>0$ (cyan and brown curves), the entropy maximum goes down because of the existence of the doublet region, but again approaches the asymptote $S = \log 2$ when the doublet region disappears at large $\lambda$'s, see Eq.~(\ref{lambda_crit}).

\section{Current through the Andreev quantum dot}

The current $I = (2e/\hbar) \, \partial E / \partial(\varphi/2)$ is defined by the operator
\begin{equation}
   {\hat I} 
   = -\frac{2e}{\hbar} \, \Gamma\sin\frac{\varphi}{2} \, [{\hat d}_\downarrow {\hat d}_\uparrow + {\rm h.c.}].
   \label{I_op}
\end{equation}
The correspondent matrix elements (in electronic Hilbert subspace)
\begin{equation}
   I_{\mu\nu}
   = -\frac{2e}{\hbar} \, \Gamma\sin\frac{\varphi}{2} \left[ \!
\begin{array}{cccc}
 0 & 0 & 0 & 1 \\
 0 & 0 & 0 & 0 \\
 0 & 0 & 0 & 0 \\
 1 & 0 & 0 & 0
\end{array}
\! \right] \!\!.
   \label{I_me}
\end{equation}

\begin{figure}[t]
  \includegraphics[width=8.4cm]{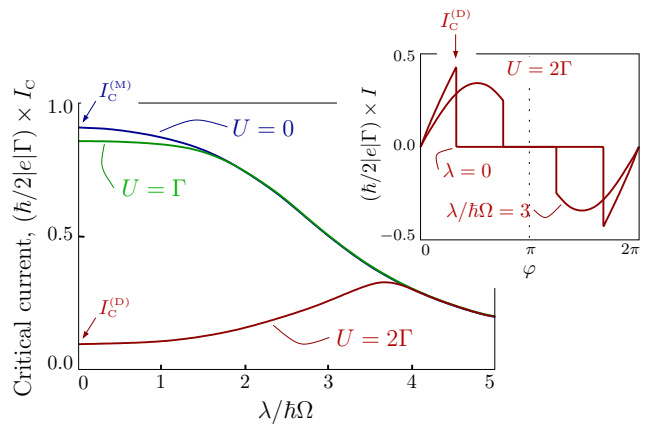}
  \caption[Critical current in Andreev quantum dot]{
  Critical current through Andreev quantum dot $I_{\rm\scriptscriptstyle C}$ as a function of electron-phonon coupling $\lambda$. Two effects compete: electron-phonon coupling dumps the current (see blue and green lines, which correspondent to $U=0$ and $U=\Gamma$) and electron-phonon coupling decreases doublet region, where $I(\varphi)=0$ (increasing of the red line and inset, at bigger Coulomb coupling $U=2\Gamma$). For large $\lambda/\hbar\Omega$ parameter the current totally suppresses. At this plot $\varepsilon_{\rm\scriptscriptstyle D} = 0.1\Gamma$.}
  \label{fig:Current}
\end{figure}

Similarly to Sec.~\ref{sec:Sensitivity} about the  sensitivity, we study the critical current $I_{\rm\scriptscriptstyle C} \equiv \underset{\varphi}{\rm max}\{I(\varphi)\}$ dependence on the coupling strength with the vibrational mode $\lambda$. The later behaves in the same way as the sensitivity except for special points at $(\pi, 0)$, see Fig.~\ref{fig:Current}.

Starting with $\lambda=0$ one finds the value of $\varphi$ which gives the maximal value of the current. It can be located at the border of the singlet-doublet region (infinitely close from the side of the singlet region) and the critical current takes the value
\begin{multline}
   I_{\rm\scriptscriptstyle C}^{\rm\scriptscriptstyle (D)} = \frac{2e}{\hbar} \frac{2}{U}
      \left\{ \frac{U^2}{4} - \Big(\varepsilon_{\rm\scriptscriptstyle D}+\frac{U}{2}\Big)^2 \right\}^{1/2} \times \\
      \left\{ \Gamma^2 - \frac{U^2}{4} + \Big(\varepsilon_{\rm\scriptscriptstyle D}+\frac{U}{2}\Big)^2 \right\}^{1/2}.
   \label{Current_c_D}
\end{multline}
Typically, if the critical current is defined by $I_{\rm\scriptscriptstyle C}^{\rm\scriptscriptstyle (D)}$, its value increases with $\lambda$, e.g., see the red line and top inset in Fig.~\ref{fig:Current}. If the critical current's $\varphi$ is located deep in singlet region its value 
\begin{multline}
   I_{\rm\scriptscriptstyle C}^{\rm\scriptscriptstyle (M)} = \frac{2e}{\hbar} \left\{
      \Gamma^2 + 2\Big(\varepsilon_{\rm\scriptscriptstyle D}+\frac{U}{2}\Big)^2 -
      \right. \\ \left.  
      2\Big|\varepsilon_{\rm\scriptscriptstyle D}+\frac{U}{2}\Big|
      \sqrt{ \Gamma^2  + \Big(\varepsilon_{\rm\scriptscriptstyle D}+\frac{U}{2}\Big)^2 }
   \right\}^{1/2}
   \label{Current_c_M}
\end{multline}
decreases with $\lambda$ (e.g., blue line in Fig.~\ref{fig:Current}). 

The existence of the doublet region transfers the system to the regime of Coulomb blockade; the electron-phonon coupling can transfer the system back to the open channel regime but simultaneously it partially suppresses the current. 

\section{Conclusion}

We considered a quantum dot with mechanical degrees of freedom which is coupled to superconducting electrodes in a Josephson junction geometry. As such a device can be used, in principle, to measure with great sensitivity the magnetic flux,~\cite{Sadovsky_Lesovik_Blatter_1} our main goal was to address the effect of the phonon coupling to this detection scheme. The superconducting gap was assumed to be larger than all relevant degrees of freedom such as the Coulomb energy and the electron-phonon coupling. In this so-called ``infinite gap limit,'' retardation effects associated with the coupling to the superconducting electrodes can be neglected, and observables can be computed using a truncated Hilbert space for the phonons.

We computed the charge on the quantum dot and determined its dependence on the phase difference in the presence of phonon coupling and Coulomb interaction. This allowed to identify regions in parameter space with the highest charge to phase sensitivity, which are relevant for flux detection. We found that nanomechanical properties significantly affect the behavior of the electron system: charge, transport, etc. In the absence of Coulomb interaction, the coupling to the vibrational mode reduces the charge sensitivity. On the other hand, when Coulomb interaction is present, it reduces (eventually completely for large coupling) the electrically insensitive doublet region due to Coulomb interaction, and in this way it increases the charge sensitivity.

Information about the entanglement properties between electron and phonon degrees of freedom was  obtained by computing the von Neumann entropy. For a fixed phase difference, and in the absence of Coulomb energy, the entropy increases with increasing phonon coupling, and eventually saturates. When plotted as a function of level position, the entropy displays a peak when the level position corresponds to the superconductor chemical potential. This peak narrows at the phase difference approaches $\pi$. When the Coulomb energy is switched on, the entropy is zero in the whole doublet region.   

%We also discussed briefly the mechanical properties of this system when several phonon modes are present.
Finally, the study of the critical current showed that for weak and moderate Coulomb energy, the current is typically reduced as the electron-phonon coupling is increased. For a larger Coulomb coupling which exceeds the dot line width, the critical current is much reduced at small electron-phonon coupling but it acquires a maximum for larger coupling strength and eventually merges with the curves corresponding to weak Coulomb interaction.
 
We acknowledge financial support by the CNRS LIA agreements with Landau Institute, the RFBR Grant No.~08-02-00767-a (IAS and GBL) and FTP ``Scientific and scientific-pedagogical personnel of innovation Russia'' in 2009--2013 (IAS).

\appendix

\section{Detailed explanation of the charge behavior
\label{app:Charge}}

\begin{figure}[t]
  \includegraphics[width=8.5cm]{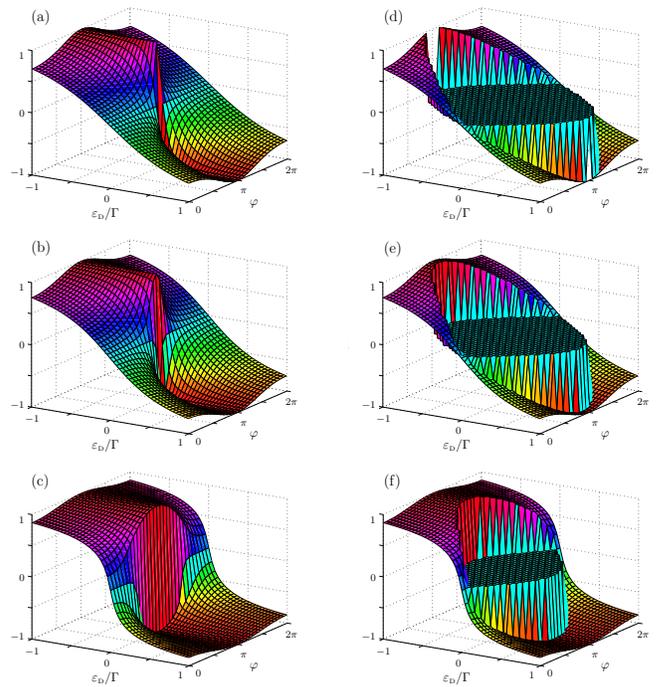}
  \caption[Charge of Andreev quantum dot as a function of $(\varphi, \varepsilon_{\rm\scriptscriptstyle D})$]{
  Charge as a function of $(\varepsilon_{\rm\scriptscriptstyle D}, \varphi)$ for zero $U=0$ [(a)-(c)] and nonzero $U=2\Gamma$ [(d)-(f)] Coulomb interaction and different $\lambda$'s. Single phonon mode with frequency $\Omega$ much smaller then tunnel resonance width $\Gamma$, $\hbar\Omega/\Gamma = 0.05$.
  (a) $\lambda=0$, the charge $Q_0$ is given by Eq.~(\ref{S_Q}). The maximal values of differential charge-to-flux sensitivity [Eq.~(\ref{Sens})] lies near point $(\varphi = \pi, \varepsilon_{\rm\scriptscriptstyle D} = 0)$. For a symmetric barrier and $U=0$ sensitivity $S(\varphi, \varepsilon_{\rm\scriptscriptstyle D})$ has the special point at $(\pi,0)$ but it disappears with any finite asymmetry, Coulomb interaction, or electron-phonon interaction.
  (b) With increasing vibration coupling $\lambda/\hbar\Omega=1.5$ the maximal charge-to-flux sensitivity decreases.
  (c) $\lambda/\hbar\Omega=3$.
  (d) Nonzero Coulomb interaction $U=2\Gamma$, no electron-phonon coupling $\lambda=0$ the largest doublet region is described by Eq.~(\ref{D_reg}).
  (e) Increased $\lambda/\hbar\Omega=1.5$ decreases the doublet region
  (f) $\lambda/\hbar\Omega=3$. 
  For a finite temperature $T$ the border of the single-doublet regions is smeared, with a width of $k_{\rm\scriptscriptstyle B}T$ in (d)-(f).
  }
  \label{fig:Charge3D}
\end{figure}

For zero Coulomb interaction $U=0$ the doublet region is absent and for $\lambda=0$ the charge is given by Eq.~(\ref{S_Q}) everywhere. For $U=\lambda=0$ the derivative $\partial Q / \partial \varphi$ has a jump at the location $(\varphi, \varepsilon_{\rm\scriptscriptstyle D}) = (\pi, 0)$ in the 2D plane $(\varphi, \varepsilon_{\rm\scriptscriptstyle D})$, see Fig.~\ref{fig:Charge3D}(a); near this point the charge-to-phase sensitivity tends to infinity. Note that for asymmetric barriers this point with an infinite slope does not exist and maxima of sensitivity are reached at four locations around $(\pi, 0)$, see Ref.~\onlinecite{Sadovsky_Lesovik_Blatter_2}. When the electron-phonon interaction is switched on this special point disappears and the maximal sensitivity is therefore suppressed by the electron-phonon interaction. This is displayed with increasing electron-phonon coupling in Figs.~\ref{fig:Charge3D}(b) and \ref{fig:Charge3D}(c). 

Next, if one now considers nonzero Coulomb interaction [Fig.~\ref{fig:Charge3D}(d)], then a ``flat'' doublet region exists: the  special point with infinite sensitivity disappears for any $U>0$ and the sensitivity in the neighborhood of this point is totally suppressed by the existence of the doublet region. In the singlet region the charge is given by Eq.~(\ref{D_Q}). 

In addition, we observe that at the boundary between the doublet and the singlet region, the charge exibits jumps and therefore the sensitivity is once again singular (for finite temperatures $T>0$ this jump is smeared) because one abruptly changes the nature of ground-state upon varying $(\varphi, \varepsilon_{\rm\scriptscriptstyle D})$. 

We next study the combination of electron interaction on the dot with the electron-phonon coupling. By increasing (from zero) the strength of the electron-phonon interaction $\lambda$, the size of the doublet region decreases. This is displayed in Figs.~\ref{fig:Charge3D}(e) and \ref{fig:Charge3D}(f) where the same electron-phonon coupling parameters are chosen as in Figs.~\ref{fig:Charge3D}(b) and \ref{fig:Charge3D}(c). The evolution of the doublet region is plotted in Fig.~\ref{fig:D_reg} for more values of the electron-phonon coupling constant. The largest loop corresponds to the smallest (zero) $\lambda$. Upon switching $\lambda$, the reduction of this region is barely noticable, it acts mostly on the level position range as it still approaches the phase values $0$ and $2\pi$. Beyond  $\lambda=0.4$, the reduction is effective in both the $(\varphi, \varepsilon_{\rm\scriptscriptstyle D})$ direction. 

There is a competition between charge repulsion effects on the dot and the presence of the electron-phonon coupling, which can be understood to be playing the role of an effective attractive, retarded, interaction. This explains the reduction of the doublet region. When comparing Figs.~\ref{fig:Charge3D}(c) and \ref{fig:Charge3D}(f) we note that the overall topology of the plots is the same except for the fact that a reduced doublet region persists at $U\neq 0$.

\end{document}